\documentclass[prl,twocolumn,superscriptaddress,floatfix,preprintnumbers,amss
ymb
,amsmath]{revtex4}
\usepackage{graphicx}
\usepackage{dcolumn}
\usepackage{bm}
\usepackage[latin1]{inputenc}
\usepackage[mathscr]{eucal}
\usepackage{epsfig}
\usepackage{rotating}

\begin{document}
\preprint{ }

\title{Shot-noise-driven escape in hysteretic Josephson junctions}

\author{J.P. Pekola}
\affiliation{Low Temperature Laboratory, Helsinki University of
Technology, P.O. Box 3500, 02015 TKK, Finland}
\author{T.E. Nieminen}
\affiliation{Low Temperature Laboratory, Helsinki University of
Technology, P.O. Box 3500, 02015 TKK, Finland}
\author{M. Meschke}
\affiliation{Low Temperature Laboratory, Helsinki University of
Technology, P.O. Box 3500, 02015 TKK, Finland}
\author{J.M. Kivioja}
\affiliation{Low Temperature Laboratory, Helsinki University of
Technology, P.O. Box 3500, 02015 TKK, Finland}
\author{A.O. Niskanen}
\affiliation{Low Temperature Laboratory, Helsinki University of
Technology, P.O. Box 3500, 02015 TKK, Finland}
\affiliation{VTT Information Technology, Microsensing, P.O. Box 1207, 02044 
VTT, Finland}
\author{J.J. Vartiainen}
\affiliation{Low Temperature Laboratory, Helsinki University of
Technology, P.O. Box 3500, 02015 TKK, Finland}


\begin{abstract}
We have measured the influence of shot noise on hysteretic Josephson 
junctions initially in macroscopic quantum tunnelling (MQT) regime. Escape 
threshold current into the resistive state decreases monotonically with 
increasing average current through the scattering conductor, which is another 
tunnel junction. Escape is predominantly determined by excitation due to the 
wide-band shot noise. This process is equivalent to thermal activation (TA) 
over the barrier at temperatures up to about four times above the critical 
temperature of the superconductor. The presented TA model is in excellent 
agreement with the experimental results.
\end{abstract}
\maketitle
Shot noise and full counting statistics (FCS) in mesoscopic conductors are 
currently intensively studied, because such measurements yield fingerprints 
of conduction mechanisms of these scatterers 
\cite{blanter,nazarov,levitov,reulet}. Yet direct measurements of the second 
and higher moments of current or voltage are typically much more difficult 
and prone to errors as compared to the measurement of DC transport 
properties. This is because, in order to extract information about 
fluctuations, one generally needs to perform high frequency measurements on 
remotely connected samples at cryogenic temperatures, which is in practise a 
non-trivial task. Therefore, detectors of noise and FCS that operate directly 
on-chip near the noise source are of great importance.  A recently proposed 
read-out device of these fluctuations is a Josephson junction (JJ) threshold 
detector \cite{tobiska,jp04}. A mesoscopic non-hysteretic Josephson junction 
in the Coulomb blockade regime is another choice 
\cite{schoelkopf,lindell,heikkila}.

\begin{figure}
\begin{center}
\includegraphics[width=0.4
\textwidth]{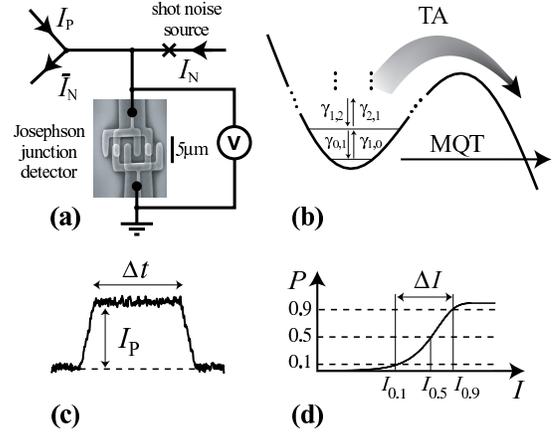}
\caption{(a) Measurement configuration. A split JJ in MQT regime measures the 
current fluctuations due to the shot noise of the scattering junction in the 
right arm. (b) Processes leading to escape from zero-voltage (supercurrent) 
state into resistive state. (c) Sketch of current pulses $I_{\rm P}$, with 
superimposed noise of $I_{\rm N}$, and (d) the associated escape 
characteristics. The central quantities relating to the pulses and escape 
histograms have been indicated in the figure.}\label{fig1}
\end{center}
\end{figure}

The dynamics of the JJ can be described as that of a phase ($\varphi$) 
particle in a tilted cosine potential, see Fig.~\ref{fig1}. Under the 
influence of equilibrium environment fluctuations at low temperatures the 
particle resides in the quantum mechanical ground state, from where the 
escape mechanism is tunnelling (MQT) through the barrier. At higher 
temperatures it assumes a nearly thermal population of the states, and the 
escape mechanism is predominantly thermal activation (TA) over the barrier 
top. Below we discuss the regime where excited states are not accessed by the 
influence of the thermal noise of the dissipative environment, which is an 
appropriate assumption based on our present measurements in the absence of 
shot noise and those of Ref.~\cite{kivioja} in the same set-up at low 
temperatures.

The influence of noisy current on escape characteristics is typically 
considered in two different limits: (i) adiabatic (quasistationary) regime, 
where fluctuations of bias current that change the tilt of the cosine 
potential are so slow and weak that the phase particle remains in the ground 
state of the metastable well, and the escape rate is dictated by MQT from 
this varying ground state \cite{martinis88,jp04}, or (ii) (nearly) resonance 
excitation limit, where most of the escape events are consequences of 
shot-noise-driven excitations of the phase particle into higher levels in the 
well. Due to the high frequency resonance character of our on-chip circuit, 
it turns out that the latter approach is more appropriate in describing the 
dynamics of the detector in this case.

In this Letter we show that a hysteretic Josephson junction in MQT regime can 
measure high frequency current fluctuations generated by a mesoscopic 
scatterer in non-equilibrium. We present a model, in which the noisy current 
excites the phase particle in a metastable, nearly parabolic well out from 
the ground state, and subsequently the particle escapes from the well. This 
occurs via thermal activation at an equivalent temperature of the phase 
particle which is determined by the competition between shot-noise 
excitation, and relaxation due to the dissipative environment. This model is 
in excellent agreement with our experimental results, and we are able to 
investigate Josephson phase dynamics up to equivalent temperatures which are 
about four times higher than the critical temperature of the superconductor. 
This is the temperature responsible for phase dynamics: the electrons and 
lattice of the superconductor are only weakly coupled to this subsystem, and 
the former two remain essentially at their base temperature.

Our measurement configuration with an electron micrograph of a split JJ 
detector is shown in Fig.~\ref{fig1}. In a typical experiment two currents, 
$I_{\rm N}$ and $I_{\rm P}$, are injected into the circuit through large 
resistances ($> 1 $ M$\Omega$) at room temperature. Current $I_{\rm N}$ is 
applied constantly and it runs through another tunnel junction at a distance 
of 120 $\mu$m from the detector. This junction plays the role of the 
shot-noise source in the circuit. The DC-component of this current, 
$\bar{I}_{\rm N}$, returns through a long (3 mm) and narrow (width 2 $\mu$m) 
superconducting line such that no DC current due to $I_{\rm N}$ passes 
through the detector. We verify this balance frequently during the 
measurements. Thus, only the fluctuations of $I_{\rm N}$ are admitted through 
the detector. We probe these fluctuations by applying repeatedly trapezoidal 
current pulses of height $I_{\rm P}$ through the detector as shown in 
Fig.~\ref{fig1}~(c). These pulses are slow, typically 10 $\mu$s - 10 ms long 
and with long transient times at the beginning and at the end of the pulse to 
secure adiabatic response to them. Typically 1000 pulses at each value of 
$I_{\rm P}$ are repeated, and the escape probability $P(I_{\rm P})$ is 
obtained as the fraction of those pulses that lead to escape from the 
supercurrent state.

Let us discuss the dynamics of a hysteretic JJ in more detail. In the common 
RCSJ-model (Resistively and Capacitively Shunted Junction model), the phase 
particle in the tilted cosine potential is either trapped and oscillating in 
one of the wells, or alternatively it runs down the potential. The latter 
regime is called resistive state. The measurable quantities, {\it i.e.}, 
current $I$ and voltage $V$, are related to $\varphi$ via the Josephson 
equations: $I = I_{\rm C}\sin \varphi$ and $2eV = \hbar \frac{d\varphi}{dt}$, 
where $I_{\rm C} = \frac{2e}{\hbar}E_{\rm J}$ is the critical current of the 
junction and $E_{\rm J}$ is its Josephson coupling energy. The quality factor 
$Q \equiv \omega _{\rm p}RC$ is assumed to be $Q\gg 1$ to assure hysteretic 
dynamics. This means that once the particle leaves the well, it runs freely 
such that the voltage across the junction is close to twice the energy gap of 
the superconductor, and it can be retrapped to the zero-voltage supercurrent 
state only when current is lowered virtually to zero. Here $R$ is the 
resistive shunt of the junction, $C$ is the parallel capacitance, and $\omega 
_{\rm p} \equiv \sqrt{8E_{\rm J}E_{\rm C}q_0}/\hbar$ is the plasma frequency, 
where $E_{\rm C}=e^2/2C$ and $q_0=\sqrt{2(1-I/I_{\rm C})}$. Quantum 
mechanically the phase particle has energy states in the nearly parabolic 
wells, with energies close to those of a harmonic oscillator: $E_n \simeq 
(n+1/2)\hbar \omega _{\rm p}$.

Due to the nearly harmonic potential, transitions between the neighbouring 
levels are dominating. The rates of these transitions, see Fig.~\ref{fig1} 
(b), are determined by the spectral density $S_{I}(\omega)=\int 
_{-\infty}^{\infty}\langle I(t)I(0)\rangle \exp(i\omega t)dt$ of the current 
noise at the corresponding level separation $\omega = \omega _{j,j-1}$ as 
$\gamma _{j,j-1} \simeq (j/2\hbar\omega _{j,j-1}C)S_{I}(-\omega _{j,j-1})$ 
and $\gamma _{j-1,j} \simeq (j/2\hbar\omega _{j,j-1}C)S_{I}(+\omega 
_{j,j-1})$ for excitation and relaxation, respectively 
\cite{schoelkopf,martinis03}. The noise spectrum has two contributions, one 
due to the equilibrium environment, $S_I^{\rm env}$, and the other due to 
shot noise, $S_I^{\rm shot}$, which add incoherently: $S_I = S_I^{\rm 
env}+S_I^{\rm shot}$. At low temperature, $k_{\rm B}T \ll \hbar \omega _{\rm 
p}$, the thermal excitation is strongly suppressed, $S_I^{\rm env} (-\omega 
_{j,j-1})\rightarrow 0$, and the relaxation is due to zero point 
fluctuations, $S_I^{\rm env}(+\omega _{j,j-1})\rightarrow 2 \hbar \omega 
_{j,j-1}{\rm Re} Y(\omega _{j,j-1})$, where $Y(\omega)$ is the admittance of 
the circuit surrounding the Josephson junction. The shot noise responsible 
for transitions in the Josephson junction can be described as $S_I^{\rm shot} 
(\pm \omega _{j,j-1}) = Fe\bar{I}_{\rm N}$. Here $\bar{I}_{\rm N}$ is the 
average current through the scatterer, and $F$ is the "Fano factor" of the 
noise source and the circuit surrounding the JJ, at the frequency 
corresponding to level separation. Factor $F$ can be calculated for a known 
experimental circuit. Here we determine it experimentally as a fit parameter. 
It would be unity in the case of Poissonian tunnel junction source 
\cite{blanter} and if all the noise current would run through the detector 
junction. Combining the results above, we have
\begin{equation} \label{rateup}
\gamma _{j,j-1} \simeq \frac{jFe\bar{I}_{\rm N}}{2\hbar\omega _{j,j-1}C}
\end{equation}
and
\begin{equation} \label{ratedown}
\gamma _{j-1,j} \simeq \frac{jFe\bar{I}_{\rm N}}{2\hbar\omega _{j,j-1}C} + 
\frac{j\omega _{j,j-1}}{Q}.
\end{equation}
The level dynamics described by Eqs.~(\ref{rateup}) and (\ref{ratedown}) can 
be described by the equivalent temperature $T^*$ and effective quality factor 
$Q^*$ by requesting $\gamma _{j,j-1}\equiv j\frac{\omega 
_{j,j-1}}{2Q^*}[\coth(\hbar \omega _{j,j-1}/2k_{\rm B}T^*)-1]$ and $\gamma 
_{j-1,j}\equiv j\frac{\omega _{j,j-1}}{2Q^*}[\coth(\hbar \omega 
_{j,j-1}/2k_{\rm B}T^*)+1]$. This yields $Q^*=Q$ and, with $\omega 
_{j,j-1}\simeq \omega _{\rm p}$
\begin{equation} \label{tstar}
k_{\rm B}T^* \simeq \frac{\hbar \omega _{\rm p}}{2 {\rm arcoth}(1+ 
QFe\bar{I}_{\rm N}/\hbar\omega _{\rm p}^2C)}.
\end{equation}
Following the standard results of the decay from a cubic metastable well, we 
can infer that thermal activation is the dominant escape mechanism provided 
$T^* > T_0 \equiv \hbar \omega _{\rm p}/2\pi k_{\rm B}$. This yields a 
condition for the validity of the TA model: $ {\rm arcoth}(1+ QFe\bar{I}_{\rm 
N}/\hbar\omega _{\rm p}^2C)/\pi < 1$. This is fulfilled at all currents 
$\bar{I}_{\rm N}$ that were employed in the experiments, as we will show 
below. With this procedure it is then straightforward to obtain the switching 
probability of the threshold detector, in the limit of many levels in the 
well, as $P(\bar{I}_{\rm P}) = 1- \exp(-\Gamma \Delta t)$. Here $\Gamma = 
\frac{\omega _{\rm p}}{2\pi}\exp(-\Delta U/k_{\rm B}T^*)$ is the standard TA 
escape rate, $\Delta t$ is the length of the current pulse, and $\Delta U = 
4\sqrt{2}E_{\rm J}(1-I_{\rm P}/I_{\rm C})^{3/2}/3$ is the height of the 
potential barrier at the particular bias point $I_{\rm P}$. The expressions 
above allow us to evaluate the position $I_{0.5}$ and the width $\Delta I = 
I_{0.9} - I_{0.1}$ of the escape threshold, where $I_x$ is defined by 
$P(I_x)\equiv x$, see Fig. \ref{fig1}~(d). These approximate results were 
compared to those obtained by full level dynamics calculation \cite{larkin} 
numerically. They were identical, although there the results depended 
sensitively on the exact values of the level positions and widths near the 
barrier top.

In the limit of large noise currents, $\bar{I}_{\rm N} \gg \frac{\hbar \omega 
_{\rm p}^2C}{QFe}$, we obtain $T^* \simeq \frac{QFe\bar{I}_{\rm N}}{2k_{\rm 
B}\omega _{\rm p}C}$, which can be written in a more familiar looking form 
$T^* \simeq \frac{Fe\bar{I}_{\rm N}}{2k_{\rm B}/R}$. This is a valid 
approximation in a wide range of experimental parameters. The expressions of 
TA escape and the current dependence of the barrier height combined with this 
approximate expression of $T^*$ finally allow us to approximate $I_{0.5}$ and 
$\Delta I$ as
\begin{equation} \label{I50}
I_{0.5}/I_{\rm C} \simeq 1 - (\frac{3}{4\sqrt{2}})^{2/3} 
[\ln(\frac{\omega_{\rm p}\Delta t}{2\pi\kappa _{0.5}})]^{2/3} \big{(} 
QF\frac{E_{\rm C}}{E_{\rm J}}\frac{\bar{I}_{\rm N}}{e\omega _{\rm 
p}}\big{)}^{2/3}
\end{equation}
and, further assuming $\omega _{\rm p}\Delta t \gg 1$ yields
\begin{equation} \label{deltaI}
\Delta I/I_{\rm C} \simeq  \frac{\ln(\kappa _{0.9})-\ln(\kappa 
_{0.1})}{[12\ln(\frac{\omega_{\rm p}\Delta t}{2\pi})]^{1/3}} 
\big{(}QF\frac{E_{\rm C}}{E_{\rm J}}\frac{\bar{I}_{\rm N}}{e\omega _{\rm 
p}}\big{)}^{2/3}.
\end{equation}
Here $\kappa _x \equiv -\ln(1-x)$.

We report on data of two samples, A and B, which were fabricated by standard 
electron beam lithography and shadow evaporation with aluminium as the 
superconductor. The samples were measured via adequately filtered signal 
lines in dilution refrigerators at bath temperatures of 30 mK - 1 K. Sample A 
had a split JJ detector and one scattering junction as in the scheme of Fig. 
\ref{fig1}. Sample B had a single JJ as a detector, and it had two scattering 
junctions located symmetrically with respect to the detector (at a distance 
of 120 $\mu$m), and with respect to {\sl two} long injection lines. In this 
sample the two scattering junctions were made intentionally very different to 
check the invariance of the results with respect to junction properties. The 
parameters of the two samples are listed in Table \ref{table}.

\begin{figure}
\begin{center}
\includegraphics[width=0.5
\textwidth]{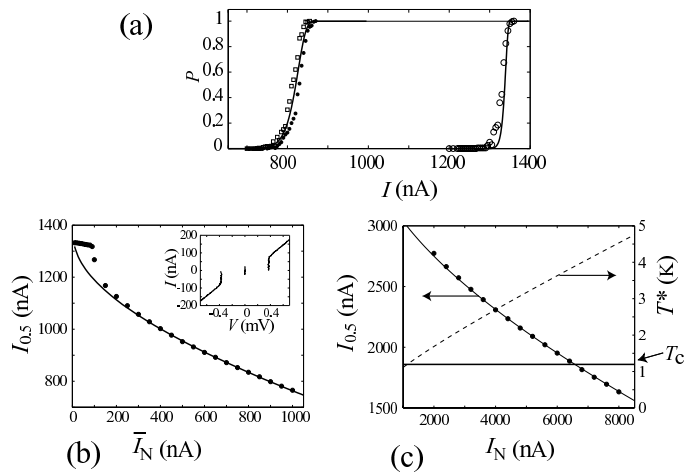}
\caption{Histogram positions under shot noise injection. (a) Escape 
histograms of Sample B when 800 $\mu$s long current pulses have been injected 
through the long injection line (open circles), and through the two noise 
sources (1 - squares, 2 - filled circles), respectively. The lines are the 
corresponding theoretical results. (b) Experimental results (circles) on the 
switching threshold current $I_{0.5}$ against the average current 
$\bar{I}_{\rm N}$ through the scatterer 2 in Sample B. Pulse length was 
$\Delta t = 800$ $\mu$s. The solid line is again the result of the used 
theoretical model. The inset shows the $IV$ curve of the noise source. (c) 
Similar data as in (b) but for Sample A. The rising curve shows the 
corresponding equivalent temperature $T^*$, which ranges from 2 to 5 K in 
this measurement.}\label{fig2}
\end{center}
\end{figure}

\begin{table}
\caption{Parameters of the two samples.}
\label{table}
\begin{tabular}{llllll}
\hline\noalign{\smallskip}
 Sample & $I_{\rm C}$ & $C$ & $I_{\rm C}$ & $C$ \\
    & detector & detector & scatterer & scatterer \\
\noalign{\smallskip}\hline\noalign{\smallskip}
A & 3.9 $\mu$A & 230 fF & 600 nA & 40 fF\\
\noalign{\smallskip}\hline
B & 1.5 $\mu$A & 230 fF & 15 nA (1) & 10 fF (1)\\
 & & & 90 nA (2)& 40 fF (2)\\
\noalign{\smallskip}\hline
\end{tabular}
\end{table}
Data in Fig.~\ref{fig2}~(a) show results of a control experiment on Sample B 
at the base temperature $T \simeq 30$ mK, where the trapezoidal pulse current 
$I$ was injected through (i) one of the long injection lines, (ii) through 
scatterer 1, and (iii) through scatterer 2. The histogram of case (i) lies at 
higher currents than those of (ii) and (iii), which in turn overlap 
practically with each other. This demonstrates that shot noise tends to push 
the threshold towards lower values of current, as predicted, {\it e.g.}, by 
Eq.~(\ref{I50}). Furthermore these data demonstrate that the noise is 
predominantly, and equally in (ii) and (iii), generated by the scattering 
tunnel junctions. The calculated lines run through the corresponding 
experimental histograms assuming MQT without any noise in (i) and
TA with $T^*$ evaluated from Eq. (\ref{tstar}) with $QF=5$ in (ii) and (iii). 
The latter value is realistic in terms of the expected $Q\simeq 10$ 
\cite{kivioja} and $F\le 1$.

Figure~\ref{fig2}~(b) shows data on Sample B at the temperature $T= 40$ mK. 
The threshold current $I_{0.5}$, {\it i.e.}, the pulse current at which the 
switching probability is 50\%, has been plotted as a function of 
$\bar{I}_{\rm N}$ through the scatterer (2). The data, shown by solid 
symbols, follow the prediction of the model presented above, again with 
$QF=5$. The initial plateau in the experimental data below $I_{\rm N} \simeq 
100$ nA arises because the scattering junction is not in the linear 
quasiparticle tunnelling regime here.
The inset of Fig.~\ref{fig2}~(b) shows the $IV$ curve of this junction.

In Fig.~\ref{fig2}~(c) we plot, in addition to similar data as in (b), the 
equivalent temperature $T^*$ in the measurement of Sample A when current 
$\bar{I}_{\rm N}$ is varied between 2 $\mu$A and 8 $\mu$A. There are two 
interesting points to note. (i) We are able to study Josephson dynamics of 
the JJ up to temperatures $T^* \simeq 5$ K, about four times above the 
critical temperature $T_{\rm C}$ of aluminium. (ii) At all currents employed 
we are well above the threshold $T_0$ of thermal activation: for this 
particular sample $T_0 \simeq 0.3$ K.
\begin{figure}
\begin{center}
\includegraphics[width=0.5
\textwidth]{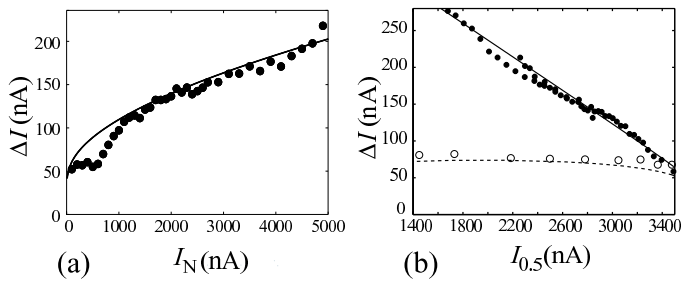}
\caption{The width of the histograms when shot noise and thermal noise are 
applied.
(a) The width of the grey zone, $\Delta I$, against the average current 
$\bar{I}_{\rm N}$ through the scatterer. The dots are the experimental 
results and the solid line is the result of the theoretical model. (b) 
$\Delta I$ vs. $I_{0.5}$ under two different experimental conditions. The 
filled circles are data with elevated shot-noise temperature $T^*$, whereas 
the open circles are with variable bath temperature $T$. The solid line is 
the result of thermal activation model with constant energy gap, and the 
dashed line takes into account the suppression of the BCS gap \cite{tinkham} 
in the measurement with increased $T$.}\label{fig3}
\end{center}
\end{figure}

Figure~\ref{fig3}~(a) shows the width of the switching threshold, $\Delta I 
$, of sample A. Again the experimental (circles) and theoretical (line) 
results are in good agreement between each other using the same fit 
parameters as in Fig.~\ref{fig2}.
In Fig.~\ref{fig3}~(b) data on switching of Sample A has been plotted in two 
different experimental conditions, in each case as $\Delta I$ vs. $I_{0.5}$. 
The open circles are data where the bath (lattice) temperature has been 
varied by heating the sample stage to several temperatures $T$ in the range 
0.03 - 1 K. The full circles are data taken at the base temperature of the 
cryostat ($T\simeq 30$ mK), but by varying the equivalent temperature $T^*$ 
injecting different levels of $\bar{I}_{\rm N}$. The line closely following 
the latter data is that originating from the thermal activation model. At the 
first sight it is surprising that the truly thermal data (open circles) fall 
far below this line and the other set of data. This is, however, accounted 
for by the fact that when increasing the bath temperature close to $T_{\rm 
C}$ of aluminium ($\sim 1.2$ K), the BCS energy gap is progressively 
diminishing, thus leading to decrease of the critical current. The line 
following closely this data set is obtained by using the very same thermal 
activation model, but by taking into account suppression of the BCS gap due 
to temperature $T$ \cite{tinkham}. This figure thus demonstrates that our 
thermal activation model is in good agreement with the shot noise data, and 
that it is indeed possible, by promoting shot noise, to study Josephson 
dynamics at super-$T_{\rm C}$ temperatures without suppressing 
superconductivity.

Finally there are a few more topics to address. We found out that the 
adiabatic models of rocking slowly the current bias of the JJ do not account 
for our observations on shot noise: the adiabatic models, involving no 
excitations, predict exponential dependence in variance-of-the-current of the 
tunnelling rate from the ground state through the barrier 
\cite{jp04,martinis88}. Our experimental results demonstrate, however, much 
weaker dependence, which we derived in this Letter. The (nearly) resonant 
model works in this case because the circuit responding to the shot noise 
resonates at frequencies, which are of the same order as the plasma frequency 
of the junction. Furthermore, the low frequency noise does not run through 
the detector junction but rather leaks through the bias line. Note that at DC 
we suppress the detector current due to $I_{\rm N}$ totally. The second issue 
is the assessment of an MQT detector as an absolute noise detector. Based on 
the invariance of the obtained results, especially in terms of results on 
different scatterers on the same sample, see Fig.~2~(a), we suggest that this 
kind of a detector could be made into an absolute on-chip detector of 
Fano-factors \cite{blanter}, and noise in general, by careful tailoring of 
the circuit surrounding the JJ and by measuring $Q$ independently.

We thank T. Heikkil\"a, T. Ojanen, E. Sonin and A. Savin for useful 
discussions, and H. Grabert and D. Esteve for their insightful comments at an 
early stage of this work. We acknowledge Academy of Finland for financial 
support.

\end{document}